\documentstyle[psfig,aps,multicol,pre]{revtex}
\newcommand{\be}{\begin{eqnarray}}
\newcommand{\ee}{\end{eqnarray}}

\newcommand{\sanda}{{\sf sanda}}
\begin{document}
\title{Critical and Near-Critical Branching Processes}
\author{Christoph Adami$^{1,2}$ and Johan Chu$^3$}
\address{$^1$Digital Life Laboratory 136-93, California Institute of
Technology, Pasadena, CA 91125\\
$^2$Jet Propulsion Laboratory MS 126-347,
California Institute of Technology, Pasadena, California 91109\\
$^3$Department of Physics, California Institute of Technology, Pasadena, CA 91125}


\draft
\maketitle
\begin{abstract}
Scale-free dynamics in physical and biological systems can arise from
a variety of causes. Here, we explore a branching process which leads
to such dynamics. We find conditions for the appearance of power laws
and study quantitatively what happens to these power laws when such
conditions are violated. From a branching process model, we predict
the behavior of two systems which seem to exhibit near scale-free
behavior---rank-frequency distributions of number of subtaxa in
biology, and abundance distributions of genotypes in an artificial life
system. In the light of these, we discuss distributions of avalanche
sizes in the Bak-Tang-Wiesenfeld sandpile model.
\end{abstract}
\begin{multicols}{2}[]
\narrowtext
\section{Introduction}

Scale-free distributions, or {\em power laws}, have been observed in a variety
of biological, chemical and physical systems. Such distributions can arise
from different underlying mechanisms, but always involve a
{\em separation} of scales, which forces the distribution to take a standard
form. Scale-free distributions are most often observed in the distribution
of sizes of events (such as the Gutenberg-Richter law~\cite{RICHTER}),
the distribution of times between events (e.g., the inter-event interval
distribution in neuronal spike trains~\cite{NEURON}), and frequencies.
An example of the latter is the well-known and ubiquitous $1/f$ noise.
Some systems are even more interesting because they seem to exhibit
self-organization or self-tuning, concomitant with
scale-free behavior as an inherent and robust property of the system,
not due to the tuning of a control parameter by the experimenter.

Two systems to which such spontaneous scale-free behavior has been
attributed are sandpile models and taxon
creation in biological systems. The former has served as the paradigm
of ``self-organized criticality''(SOC)~\cite{BTW},
while the latter,
manifested in the form of near power law shapes of rank-abundance curves,
has been advanced as evidence of a fractal geometry of
evolution~\cite{BURLANDO}.

A much simpler system where power laws are observed is the random
walk~\cite{SPITZER}.
For example, the waiting times $t$ for first return to zero of the
simple random walk in one dimension (starting at $x=0$, at each time
step, $x(t+1)=x(t) \pm 1$ with equal probability)
have a probability distribution $\sim t^{-3/2}$.
Closely related to random walks, branching processes~\cite{HARRIS}
can also create power law distributions. They have been
used to model the dynamics of many systems in a wide variety of disciplines,
including demography, genetics, ecology, physiology, chemistry,
nuclear physics, and astrophysics.
Here, we use a branching process to model the creation and growth
of evolutionary taxa, and discuss its application to avalanches in SOC
sandpile models.

In Section II, we examine the properties of the {\it Galton-Watson} process.
We find that
this process can generate power laws by appropriate tuning of a control
parameter, and examine the dynamics of the system both at the critical
point and away from it.  In Section III,
we apply this branching process model to
the taxonomic rank-frequency abundance patterns of evolution, and discuss the
universality of their underlying dynamics.
Finally, in Section IV, we discuss the implications of our work,
including a discussion of the order and control parameters for the
branching process and its applications, and suggest further questions.

\section {The Branching Process}

The Galton-Watson branching process was first introduced in 1874 to explain
the disappearance of family names among the British peerage~\cite{WATSON}.
It is the first branching process in the literature, and also
one of the simplest. Consider an organism that replicates. The number of
replicants ({\em daughters}) it spawns is determined probabilistically, with
$p_i$ ($i=0,1,2$...) being the probability of having $i$ daughters.
Each daughter replicates (with the same $p_i$ as the original organism) and
the daughter's daughters replicate and so on.
We are interested in the rank-frequency probability distribution
$P(n)$ of the total number of
organisms descended from this organism plus $1$ (for the original organism),
i.e., the historical size of the ``colony'' the
ancestral replicant has given rise to.
Note that this is equivalent to asking for the probability distribution
of the length of a random walk starting from 1 and returning to 0 with
step sizes given by $P(\Delta n) = p_{i-1}$ ($i=0,1,2$...)~\cite{ABH}.

The abundance distribution $P(n)$ can be found by defining a generating
function
\be
F(s)=\sum_{i=1}^{\infty} P(i) s^i.
\ee
This function satisfies the relationship
\be
F(s) = s \sum_{i=0}^{\infty} p_i [F(s)]^i,
\ee
from which each $P(n)$ can be determined by equating coefficients of
the same order in $s$~\cite{HARRIS}.
This result can also be written as
\be
P(n) = \frac{1}{n} Q(n,n-1) \qquad (k \geq 1),
\ee
where $Q(i,j)$ is defined as the probability that $j$ organisms will give
birth to a total of $i$ true daughters~\cite{SPITZER}.
However, these approaches are not numerically efficient, as
the calculation of $P(n)$ for each new value of $n$ requires re-calculation
of each term in the result.

For our present purposes, we approach the problem in a different manner. Let
$P_{k|j}$ be the probability that given $j$ original organisms,
we end up with a total of $k$ organisms after all organisms have finished
replicating. Obviously,
\be
P_{k|j} = 0 \qquad (k < j), \label{START}
\ee
since it is impossible to have less total organisms than one starts out with,
and
\be
P_{1|1} = p_0,
\ee
i.e., the probability for one organism to have no daughters. A little less
obviously,
\be
P_{k|1} & = & \sum_{j=1}^{k-1} p_j P_{(k-1)|j}, \label{PROP} \\
P_{k|j} & = & \sum_{i=1}^{k-1} P_{i|1} P_{i|(j-1)} \qquad (j \geq k > 1).
         \label{END}
\ee
These equations allow us to use dynamic programming techniques to calculate
$P(n)$ ($ = P_{n|1}$), significantly reducing the computational time required.
Also, from Eq. (\ref{PROP}), we can write
\be
 \frac{P_{n|1}}{P_{(n-1)|1}} = p_1 + p_2 \frac{P_{(n-1)|2}}{P_{(n-1)|1}}
  + p_3 \frac{P_{(n-1)|3}}{P_{(n-1)|1}} + \hbox{... }.
\ee
Since, for $n \rightarrow \infty$, $P_{n|j}$ is uniformly decreasing, we
see
\be
\frac{P(n)}{P(n-1)} = \frac{P_{n|1}}{P_{(n-1)|1}} \rightarrow C \hbox{ }
\hbox{ as } n \rightarrow \infty, \quad (C \leq 1)
\ee
where $C$ is a constant. $C$ indicates the asymptotic behavior of
$P(n)$ as $n \rightarrow \infty$. If $C < 1$, the probability distribution is
asymptotically exponential, while for $C = 1$, the probability distribution
is a power law with exponent $-3/2$.

\begin{figure}[h]
\centerline{\psfig{figure=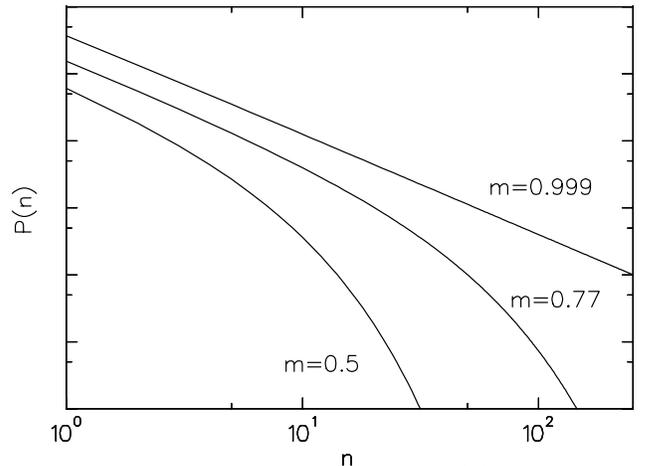,width=3.25in,angle=90}}
\caption{Predicted abundance patterns $P(n)$ of the branching model with
different values of $m$. The curves have been individually re-scaled.}
\label{THEORYF}
\end{figure}

Let us now examine the behavior of $P(n)$ when $n \lesssim 10^4$, the
more relevant case in the examples to follow.  Using
Eqs. (\ref{START})-(\ref{END}), we can numerically calculate $P(n)$
for different sets of $p_i$. We define $m$ as the expected number of
daughters per organism, given a set of probabilities $p_i$; \be m =
\sum_i i \cdot p_i.  \ee We see that the branching rate $m$ (the {\em
control parameter}) is a good indicator of the shape of the
probability curve (Fig.~\ref{THEORYF}).  When $m$ is close to $1$, the
distribution is nearly a power law, and the further $m$ diverges from
$1$, the further the curve diverges from a power law towards an
exponential.  When $m = 1/2$, the curve is completely exponential. For
a population of organisms, $m$ is a measure of the tendency for new
generations to grow, or shrink, in number. A value of $m > 1$
indicates a growing generation size, which implies that there will, on
average, be no generation with no daughters, and that the expected
number of total organisms is infinite.  Conversely, $m < 1$ indicates
a shrinking population size: There will be a final generation with no
daughters, and the expected number of organisms is finite. When $m=1$,
the system is in between the two regimes (the system is said to be
``critical''), and only then is a power law distribution found. In
general, the branching rate is determined by the ratio of the rate of
introduction of competitors $R_c$ to the intrinsic rate of growth of
existing assemblages $R_p$ via
\be
m = (1+\frac{R_c}{R_p})^{-1}\;,
\ee
as can be shown by assuming stationarity. As this ratio
goes to $0$, $m \rightarrow 1$ and the system becomes critical.

In the following section, we explore systems where the ``organisms''
are individual members of species or taxa in a taxonomic tree, and $m$
is the average number of exact copies an individual makes of itself,
or the average number of new taxa of the same supertaxon a taxon
spawns, respectively. The same thinking can be applied to tumbling
sites in a sandpile model, where $m$ would stand for the average
number of new tumbles directly caused by a tumbling site.

\section {Applications}
\subsection{Neutral Model}

\begin{figure}[h]
\centerline{\psfig{figure=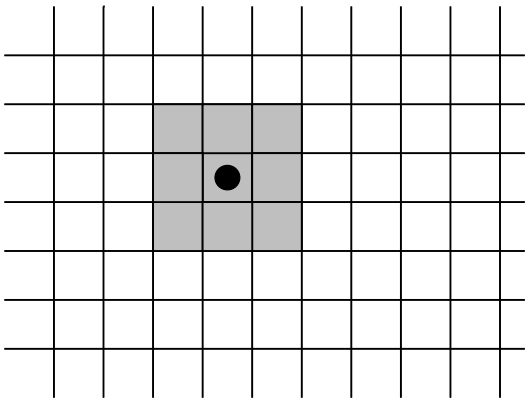,width=3.25in,angle=-90}}
\caption{Neutral model grid. The organisms live on an Euclidean grid, one
organism to a site. When an organism replicates, its daughter replaces the
oldest organism in the $9$-site neighborhood. (If the organism marked by
a black dot replicates, its daughter replaces one of the organisms at a
gray site.)}
\label{GRIDF}
\end{figure}

\begin{figure}[h]
\vskip 0.25cm
\centerline{\psfig{figure=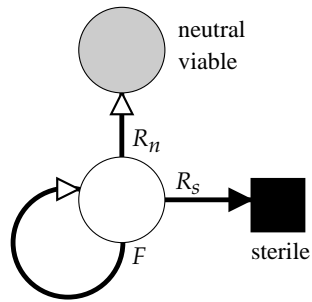,width=3.25in,angle=-90}}
\caption{Neutral replications and mutations. An organism's daughter
is of the same genotype as the organism with probability $F$, it is
of a new, viable genotype with probability $R_n$, and it is sterile
with probability $R_s$ such that $F+R_n+R_s=1$.}
\label{LOOPF}
\end{figure}

We first present a simple simulation to test our analysis and lay the
groundwork for the exploration of more complicated systems. Consider
a population of organisms on a finite
two-dimensional Euclidean lattice, one organism
to a grid square. Each organism can be {\em viable} or {\em sterile}.
All viable organisms replicate approximately every $\tau$ time steps (there
is a small random component to each individual's replication time to avoid
synchronization effects), while sterile organisms do not replicate.
When an organism replicates, its daughter replaces the oldest organism in the
parent's $9$-site neighborhood (Fig.~\ref{GRIDF}).
We define the {\em fidelity} $F$
as the probability that the organism will create a daughter of the same
type as itself and the corresponding {\em genomic mutation rate}
$R$ ($= 1-F$) at which
it creates copies different from itself. The genomic
mutation rate is actually the sum of two rates,
a probability $R_n$ for the daughter to be viable but to be of a new
{\em genotype},
different from that of the parent ({\em neutrality rate}), and a
probability $R_s$ of the daughter being sterile. Of course, $R_n+R_s = R$.
Note that all viable mutant daughters still share the same replication
time $\tau$---all mutations are neutral (See Fig.~\ref{LOOPF}). Such a
system gives rise to abundance distributions of power law and near-power
law type, which can be predicted as follows.

\begin{figure}[h]
\centerline{\psfig{figure=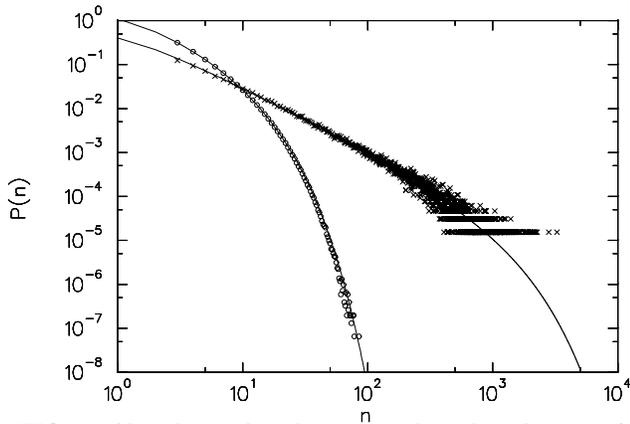,width=3.25in,angle=90}}
\caption{Abundance distributions and predicted curves for two neutral model
runs. The run shown by circles ($\sim1.5$ million data points)
had a grid size of $3000\times3000$, $F=0.5$, and $R_n=0.5$, while the
one represented by crosses ($\sim0.6$ million data points) had a grid size
of $100\times100$, $F=0.2$, and $R_n=0.1$.
}
\label{NEUTRALF}
\end{figure}

The total number of organisms is determined by the size of the grid.
We write equilibrium conditions for the total number of organisms $\rho_A$,
and for the total number of viable organisms $\rho_V$,
\be
\Delta \rho_A & \sim & a \rho_V - \rho_A = 0, \label{EQ1} \\
\Delta \rho_V & \sim & v \rho_V - \rho_V = 0,
\ee
where $a$ is the average number of daughters (viable and sterile) a viable
organism has, and $v$ is the average number of viable daughters a viable
organism has. Introducing $m$---the average number of true daughters
(daughters which share the parent's genotype) for a viable organism---we
see that
\be
v = \frac{F+R_n}{F} m = (F+R_n) a. \label{EQ3}
\ee
From Eqs. (\ref{EQ1})-(\ref{EQ3}), we obtain steady state solutions for
$a$ and $m$,
\be
a & = & \frac{F^{-1}}{1+\frac{R_n}{F}}, \label{A} \\
m & = & \frac{1}{1+\frac{R_n}{F}}. \label{M}
\ee
Using the branching process model, we can predict the abundance curve from
the values of $a$ and $m$ (or conversely, $F$ and $R_n$).
Fig.~\ref{NEUTRALF} shows abundance data
for two neutral model runs with differing values of $R_n$ (and consequently
$m$), along with predicted distributions (which use only
$R_n$ and $F$ as parameters) based on the branching model.
Although the distribution patterns are very different, both are fit
extremely well by the branching process model's predicted curves.
In Eq. (\ref{M}), note that $R_n$ is the rate of influx
of new genotypes (and therefore new competitors for space), while $F$ is the
rate of growth of existing genotypes. The value of $m$ is determined by
the ratio of these two rates.  Unless the total number of creatures
is increasing, $m \leq 1$ ($m=1$ if and only if $R_n \rightarrow 0$ and
new competing genotypes are introduced at a vanishing rate).

\subsection{Artificial Life}

\begin{figure}[h]
\centerline{\psfig{figure=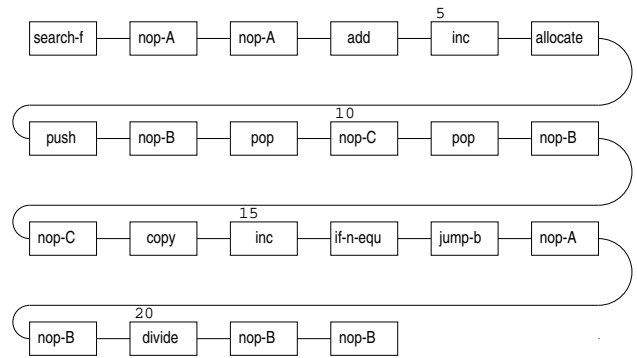,width=3.25in,angle=-90}}
\vskip 0.25cm
\caption{Example \sanda\ genotype. {\sf Sanda} organisms have genotypes
which are strings of \sanda\ code. The string shown above replicates
by: searching forward ({\sf instruction 1}) for the complement of the template
{\sf nop-A nop-A} ({\sf 2-3}), which is {\sf nop-B nop-B} ({\sf 21-22}),
manipulating this value in an internal register to find the genotype length
({\sf 4-5}), allocating enough memory to store code of genotype length
({\sf 6}), setting registers to prepare for copying ({\sf 7-11}), copying
the instructions one at a time ({\sf 12-19}) until all instructions have
been copied ({\sf 15-16}), and replicating ({\sf 20})---placing the daughter
in its own grid site. Execution restarts at the beginning of the genotype when
the end of the genotype is reached, and continues until the organism is
replaced by the newly replicated daughter of another organism (or its own
daughter). The {\sf copy} command ({\sf 14} in this particular genotype) fails
and writes a random instruction with probability $\gamma$.}
\label{SANDACF}
\end{figure}

Our next system is the artificial life system \sanda~\cite{SANDA},
an example of environments which host digital organisms~\cite{CABOOK}.
In this system, while the organisms occupy a two-dimensional grid as in
the neutral model detailed above, the organisms are no longer simple,
and instead each has a complex genotype consisting of a string of assembly
language-like instructions (Fig.~\ref{SANDACF}).
Each organism independently executes the instructions of its
genotype, and this genotype determines the organism's replication time
$\tau$. Unlike the neutral model,
the system allows non-neutral mutations which lead to new genotypes with
both lower and higher replication times than the parent.

The system and the instructions are designed so that the organisms
can self-replicate by executing certain sequences of instructions.
The replication time of an organism is not a predetermined constant,
rather it is determined by the genotype of the organism: Organisms can
replicate faster or slower than other competing organisms with different
genotypes.  For an organism to successfully replicate,
its genotype must contain information which allows the organism to
allocate temporary space (memory) for its daughter, replicate its genotype (one
instruction at a time)
into this temporary space, and then to divide, placing its daughter in
a grid site of its own (Fig.~\ref{SANDACF}).
As in the neutral model, on division,
the daughter replaces the oldest organism in its parent's $9$-site neighborhood.

Organisms, depending on their genotype, may not be able to replicate
(may be sterile) or may only be able to replicate imperfectly (resulting
in no true
daughters). Also, the {\sf copy} instruction, which the organisms must
use to copy instructions from their own code into that of their nascent
daughters, has a probability of failure ({\em copy mutation rate}),
which can be set by the experimenter.
When the {\sf copy} instruction
fails, an instruction is randomly chosen from all the instructions available
to the organisms (the {\em instruction set}) and written in the string location
copied to. Copy mutations also lead to non-true daughters.
The instruction set is robust; copy errors (mutations) induced during the
replication of viable organisms have a non-vanishing probability of creating
viable new organisms and genotypes. Indeed, by selecting for certain traits
(such as the ability to perform binary logical operations) by
increasing the relative speed at which instructions are executed in organisms
which carry these traits, the system can
be forced to {\em evolve} and find novel genotypes which contain more
information (and less entropy) than their ancestors. Even without this
external selection, the system evolves organisms (and genotypes) which
replicate more efficiently in less executed instructions.

As a result of this evolution, the fidelity and neutral mutation rate
are not fixed, but can vary with the length of an organism's genome
and the instructions contained therein. Also, new genotypes formed by
beneficial mutations that allow faster replication than previously
existing genotypes will have (on average) an increasing number of
organisms---$m>1$---until the new, faster replicating genotypes fill
up a sizable portion of the grid. All these factors combine to make
predicting the abundance distributions for \sanda\ much harder than
for the neutral model.

Indeed, rather than being constant during the course of a \sanda\ experiment,
$R_n$ and $F$ will vary unpredictably
as the population of organisms occupies different areas in genotypic
phase space. Certain genotypes may be {\em brittle}, allowing very few
mutations that result in new viable genotypes. The length of the
organisms may change, changing both the genomic mutation rate and the
neutrality rate. Genotypes exist which make systematic errors when copying,
which decreases the fidelity. In short, the dynamics of these digital organisms
are complex and messy, much like those of their biochemical brethren.
These variations are observed at the
same time across different organisms in the population, and are also
observed with the progression of time.
Still, we attempt to predict the abundance distributions by approximating
the ratio of neutral mutations to true copies by the {\em observed} ratio
of viable genotypes to total number of viable organisms ever created:
\be
\frac{R_n}{F} \simeq \frac{N_g}{N_v},
\ee
where $N_g$ is the total number of viable genotypes observed during a
\sanda\ run and $N_v$ is the total number of viable organisms. This relation
should hold approximately under equilibrium conditions. Then,
Eq.~(\ref{M}) becomes
\be
m \simeq (1+\frac{N_g}{N_v})^{-1}, \label{M2}
\ee
and from Eq. (\ref{A})
\be
a = \frac{m}{F}.
\ee
The fidelity $F$ is inferred from the average length $l$ of
genotypes during a run and the (externally enforced) per-instruction
copy mutation rate $\gamma$, $F = (1-\gamma)^l$. Because we estimate
$m$ and $a$ from macroscopic observables averaged over the length of a run,
we expect some error in our results due to the shifting dynamics of the
evolution of genotypes as the system moves in genotypic phase space.

\begin{figure}[h]
\centerline{\psfig{figure=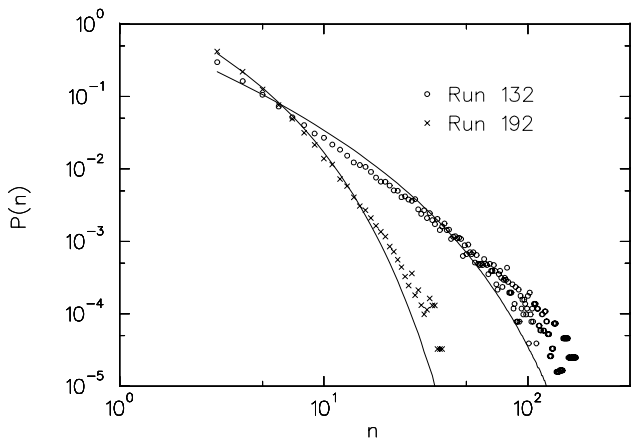,width=3.25in,angle=90}}
\caption{Abundance data from two \sanda\ runs with predicted abundance curves.
Both runs were started with the
same initial genotype for all organisms, the same per-instruction
copy mutation rate ($\gamma$), and the same grid size ($100 \times 100$).
Run 192's genotypes
evolved into a regime of genotypic phase space with longer average length,
and therefore lower fidelity $F$ and higher neutrality $R_n$, than
Run 132, resulting
in the differences in the abundance distributions. The predicted curves
were generated by approximating a representative value of $R_n/F$ from the
ratio of
the number of viable genotypes to the number of viable organisms observed
over the run. The data was binned using the template threshold
method with $T=1$ (see Appendix).}
\label{SANDAF}
\end{figure}

The abundance data from
two different \sanda\ runs are shown in Fig.~\ref{SANDAF} with the predicted
abundance curves. The two runs shared the same grid size and per-instruction
copy mutation rate, and were started with the same initial
genotypes, but the runs evolved into different regions
of genotypic phase space and consequently had significantly different
statistics.
Considering the many
gross approximations we have made, the agreement between our prediction
and the experimental data is surprisingly good (especially as no fitting
is involved).
{\sf Sanda} is most closely related to an asexually replicating biological
population, such as colonies of certain types of bacteria occupying a single
niche. The genotype abundance distributions measured in \sanda\ are analogous
to the species or subspecies abundance distributions of its biological
counterparts.
In general, species abundance distributions are complicated by the
effects of sexual
reproduction, and of the localized and variable influences of other species
and the environment on species abundances. However, we believe the branching
model---used judiciously---can be helpful in the study of such
distributions as well.

\subsection{Evolution}

Rank-abundance distributions at taxonomic levels higher than species
(e.g., the distribution of the number of families per order)
are simpler to model than species abundance distributions, as the
effects of the complications noted above are weak or nonexistent.
We find that the available data is well fit by assuming no
direct interaction or fitness difference between taxa~\cite{JCCABIO}.
The shapes of rank-frequency distributions of taxonomic and evolutionary
assemblages found in nature are surprisingly uniform. Indeed, Burlando has
speculated that all higher-order taxonomic rank-frequency distributions
follow power laws stemming from underlying fractal
dynamics~\cite{BURLANDO}. We believe this conclusion is hasty:
The divergence of the distributions from power law can be observed
by applying appropriate binning methods to the data. (See Appendix.)
Yule~\cite{YULE} attempted a
branching process model explanation of these distributions, and claimed
that divergence from power law of rank-abundance patterns was transient
and indicated a finite time since the creation of the evolutionary
assemblage.
Our model indicates that this is not generally the case. We find that the
divergence
from power law is not a result of disequilibration, but is an inherent property
of the evolutionary assemblage under consideration and that this
divergence provides insight into microscopic properties of the assemblage
(e.g., the rate of innovation).

Say, for example, that we are interested in the rank-frequency distribution
of the number of families in each order for fossil marine animal orders.
We assume that all new families and orders in this assemblage originate
from mutations in extant families. Then, we can define rates of successful
mutation $R_f$ for mutations which create new families in the same order
as the original family, and $R_o$ for mutations which create an entirely
new order. In this case, unlike the cases treated
above, we approximate $a \rightarrow \infty$; many individual
births and mutations occur, but the proportion that are family- or
order-forming is minuscule. Finally, assuming a quasi-steady state
(the total numbers of orders and families vary slowly~\cite{RAUP}), we rewrite
Eq. (\ref{M}),
\be
m & \simeq & (1+\frac{R_o}{R_f})^{-1} \\
& \simeq & (1+\frac{N_o}{N_f})^{-1},
\ee
in terms of $N_o$ and $N_f$, the total numbers of orders and families
respectively.
As in the previous systems studied, $R_o$ is the rate of creation of
new---competing---orders,
while $R_f$ is the rate of growth of existing orders, and
$m$ is determined by their ratio.

\begin{figure}[h]
\centerline{\psfig{figure=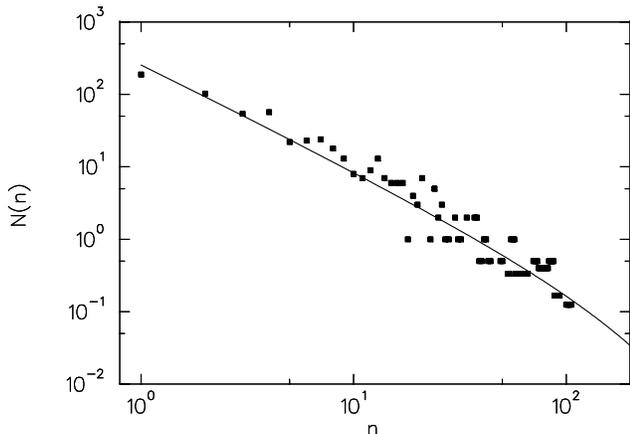,width=3.25in,angle=90}}
\caption{The rank-frequency distribution of fossil marine animal orders
(squares)~\protect\cite{SEPKOSKI}
and the predicted abundance curve (line). The
predicted curve was generated---with no free parameters---by approximating
$R_n/F$ by $N_o/N_f = 0.115$. The empirical distribution agrees with the
predicted curve with significance $0.12$ using the Kolmogorov-Smirnov
test~\protect\cite{JCCABIO}.
The fossil data is shown binned using the template threshold binning method
explained in the Appendix with $T=1$.}
\label{FOSSILF}
\end{figure}

Data for the abundance distribution of number of families in
fossil marine animal orders~\cite{SEPKOSKI} are shown in Fig.~\ref{FOSSILF}.
We obtained values for $N_o$ and $N_f$ directly from the fossil data to generate
the predicted curve with {\em no free parameters}.
The agreement is very good, much better than that for the \sanda\ runs where
evolutionary parameters such as the fidelity $F$ and the neutrality $R_n$
were constantly changing. Comparing $m$ and the
resultant abundance curves with those obtained above for the rank-abundance
distribution of \sanda\ genotypes leads us to the expected conclusion
that the probability of creation of a new genotype in \sanda\ per birth
is much higher than the probability of a new family creating an order
in natural evolution.
Indeed, a wide variety of higher-order taxonomic assemblages have
abundance distributions consistent with $m$ near 1~\cite{BURLANDO}.
We believe this is a robust result of the evolutionary process. Low values of
$m$ may not be observed for large taxon assemblages for several reasons. A small
value of $m$ implies either a small number of individuals in the assemblage,
or a very specialized niche with a very low rate of taxon formation. A low
number of individuals would lead to a low probability of the taxon being
discovered and cataloged by biologists. A small number of individuals
and taxa would result in an
assemblage with too few taxa to give us a clear statistical picture.
Also, since such an assemblage would have a small population, be incapable
of further adaptation, or both, we expect it would be more susceptible
to competition and environmental effects leading to early extinction.

\subsection{Sandpile Models}
It was originally suggested that the self-organization observed in the
sandpile model of Bak, Tang and Wiesenfeld (BTW)~\cite{BTW} (and the
power laws it displayed) was an inherent property of the system, while
it now seems established that the system is actually tuned by waiting
until avalanches are over before dropping new grains---this is
equivalent to allowing non-local
interactions~\cite{SORNETTE,ZAPPERI2}. The same conclusion is reached
when using a branching process to describe the avalanche dynamics.
Branching processes have been applied to sandpile models as early as
1988 \cite{alstrom88} (see also,
\cite{theiler93,christensen93,pelayo94,lauritsen96,ZAPPERI}). Using a
mean-field approach in higher dimensions ($d \gtrsim 4$), power law
distributions for the size of avalanches $s(n)$ can be obtained
analytically, and critical exponents can be calculated exactly to
reveal $s(n) \sim n^{-3/2}$~\cite{ZAPPERI} in the limit of
infinitesimally small driving.  This is supported by numerical
simulations.  However, for lower dimensions, sandpiles will
``interfere'' with themselves, and a smaller exponent is
found. Attempts to calculate the effects of this ``final-state''
interaction through renormalization have as yet not been completely
successful \cite{lubeck98}. Still, the phenomenon of ``violations'' of
power-law behavior due to $m<1$ (non-critical branching process) can
be seen there as well.

\section{Discussion}

The Galton-Watson branching process generates power law distributions
when its control parameter $m = 1$. In all the systems we have
examined above, \be m = (1+\frac{R_c}{R_p})^{-1} \ee is determined by
the ratio of the rate of introduction of competitors $R_c$ to the
intrinsic rate of growth of existing assemblages $R_p$. As this ratio
goes to $0$, $m \rightarrow 1$ and the system becomes critical.

\begin{figure}[h]
\centerline{\psfig{figure=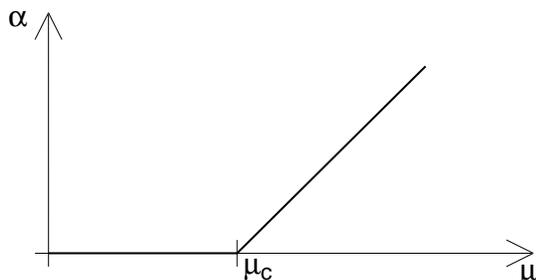,width=3.25in,angle=-90}}
\caption{The order parameter $\alpha$ as a function of the control parameter
$\mu$. For $\mu$ below $\mu_c$, the order parameter is $0$---organisms
(or events) in the system spawn greater and greater number of daughter
organisms (events), and there is exponential growth. For $\mu > \mu_c$,
competition from newly created organisms (events) stops abundances from
growing without bound. $\mu=\mu_c$ marks the critical point
where abundances can grow to infinity, but do not show exponential growth,
and power law distributions arise.}
\label{ORDERF}
\end{figure}

This relation can be translated into the standard relation between an
{\em order parameter}
\be
\alpha = \frac{R_c}{R_p}
\ee
and a new form for the control parameter
\be
\mu = m^{-1}.
\ee
Writing $\alpha$ in terms of $\mu$,
\[ \alpha = \left\{ \begin{array} {c@{\qquad}l}
(\mu - \mu_c)^\beta & (\mu > \mu_c), \\
0 & (\mu \leq \mu_c),
\end{array}
\right.
\]
where $\mu_c=1$ and $\beta=1$ (Fig.~\ref{ORDERF}). The order parameter represents
the rate at which competition is introduced in the system (the strength of
selection). A value of the control parameter $\mu < \mu_c$ implies a system with no
competition and no selection---an exponentially growing population. Values of $\mu$
higher than $\mu_c$ indicate that new competition is always being introduced and
that all existing species or avalanches must eventually die out. When $\mu=\mu_c$,
competition is introduced at a vanishingly small rate, and we find the critical
situation where separation of scales occurs.

For sandpile models, this $\alpha$ is arbitrarily set close to $0$ by using
large lattice sizes (reducing dissipation) and waiting for avalanches to
finish before introducing new perturbations (resulting in an infinitesimal
driving rate and a diverging diffusion coefficient).
For the biological and biologically-inspired systems we have considered, the
control parameter is not set arbitrarily to a critical value. However, the dynamics
of the evolutionary process, in which it is much harder to effect large jumps in
fitness and function than it is to effect small ones, lead to naturally observed
values of $\alpha$ being small, especially for higher taxonomic orders. The
dynamics of evolution act, robustly, to keep $\mu$ near $\mu_c$.  This in turn
leads to a near power law pattern for rank-frequency distributions.

We have shown that the apparent power laws of avalanches in species-abundance
distributions in artificial life systems, as well as rank-abundance distributions
in taxonomy can be explained by modeling the dynamics of the underlying system with
a simple branching process. This branching process model successfully predicts,
with no free parameters, the observed abundance distributions---including their
divergence from power law.

A branching process approach may allow the deduction of
the microscopic parameters of the system directly from the macroscopic
abundance distribution.  We find that we can identify a control
parameter---the average number of new events an event directly spawns,
and an order parameter--- the rate of introduction of competing events
into the system, and that these are related in a form familiar from
second order phase transitions in statistical physics.

\acknowledgements We are grateful to the late Prof.\ J. J.\ Sepkoski for kindly
sending us his amended data set for fossil marine animal families, as
well as an anonymous referee for insightful comments.  J. C. thanks
M. C. Cross for continued support and discussions.  Access to the
Intel Paragon XP/S was provided by the Center of Advanced Computing
Research at the California Institute of Technology.  This research was
supported by the NSF under contract Nos.\ PHY-9723972 and DEB-9981397.
Part of this work was carried out at the Jet Propulsion Laboratory,
California Institute of Technology, under a contract with the National
Aeronautics and Space Administration.
\appendix
\section{Binning Methods}
\begin{figure}[h]
\centerline{\psfig{figure=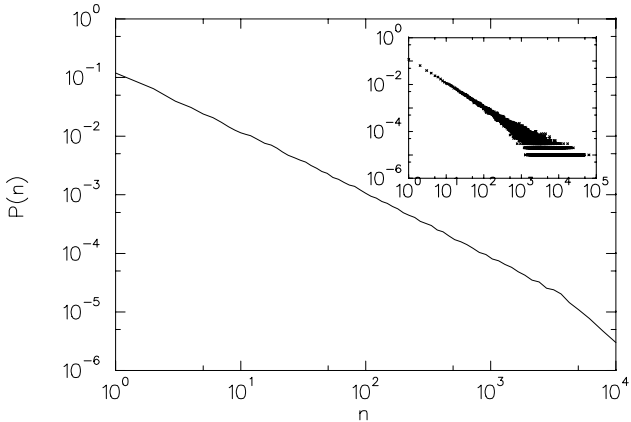,width=3.25in,angle=90}}
\vskip 0.25cm
\caption{Binned avalanche size distribution for the BTW sandpile in
the limit of infinitesimally slow driving (the standard BTW protocol).
The inset shows avalanche size distribution data after 100,000
avalanches.  The main panel shows the same data binned using the data
threshold method with $T=1000$. That this binning method accurately reproduces
the function this data is drawn from can be seen by
comparing to the data set of 16 million avalanches (Fig.~\ref{SOCF}),
which shows no discernible differences between the predictions made by
binning and the conclusions given by more data.}
\label{BINF}
\end{figure}

When dealing with event distributions best plotted on
single log or double log scales
(such as exponential and power law distributions),
care must be taken in the proper binning of the experimental data.
Say we are interested in the probability distribution $P(n)$ of an
event distribution over positive integer values of $n$.
We conduct $N$ trials, resulting in a data set $Q(n)$ of number of events
observed for every $n$ value.
For ranges of $n$ where the expected or observed
number of events for each $n$ is much higher than 1, normally no binning
is required. However, for ranges of $n$ where $Q(n)$ or $P(n)$ is
small, binning is necessary to produce both statistically significant
data points, and intuitively correct graphical representations.
A constant bin size has several drawbacks: One must guess and choose an
intermediate bin size to serve across a broad range of parameter space,
and the shape and slopes of the curve (even in a double log plot) are
distorted~\cite{CABOOK}.  These disadvantages can be overcome by using a
variable bin size. However, choosing bin sizes for variable binning is
time-consuming and arbitrary---different choices will lead to different
conclusions. We propose two related methods of systematically determining
appropriate variable bin sizes. Both methods lead to binned data which
help in visualizing the underlying distribution (slopes and shapes are
conserved).

For the first method (the {\em Data Threshold Method}),
we start by selecting a threshold value $T$.
Starting from
$n=1$ and proceeding to higher values, no binning is done until a value of $n$
is found for which $Q(n) < T$. When such a value $n_s$ is found, subsequent
$Q(n)$ values are added to this amount until the sum of these values is
greater than the threshold value,
\be
\sum_{n=n_s}^{n_l} Q(n) > T.
\ee
We then have a bin size $(n_l-n_s+1)$, with value $\sum_{n=n_s}^{n_l} Q(n)$.
When plotting, it is convenient to plot this as a single point at the
midpoint of $[n_s,n_l]$, with an averaged value,
\be
\left(\frac{n_s+n_l}{2}, \frac{\sum_{n=n_s}^{n_l} Q(n)}{n_l-n_s+1}\right).
\ee
This yields a graphical representation with little distortion and good
predictive power (Fig.~\ref{BINF}). This binning procedure is continued
until no more data remains to be binned.

\begin{figure}[h]
\centerline{\psfig{figure=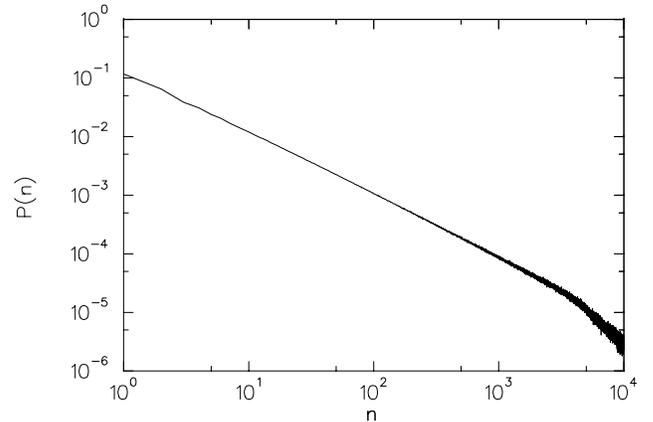,width=3.25in,angle=90}}
\vskip 0.25cm
\caption{Avalanche size distribution in the $2$-$d$
BTW sandpile model with infinitesimal driving rate (16 million avalanches).}
\label{SOCF}
\end{figure}

The second binning method (the {\em Template Threshold Method}),
uses a predicted probability distribution
$P(n)$, or a reasonable surrogate. Again, we define a threshold value
for fitting $T$. However, in this case, the bin sizes are determined by
comparing values of the {\em expected distribution}
\be
E(n)=P(n) \times N
\ee
to $T$. Starting from
$n=1$ and proceeding to higher values, no binning is done until a value of $n$
is found for which $E(n) < T$. When such a value $n_s$ is found, subsequent
$E(n)$ values are added to this amount until the sum of these values is
greater than the threshold value,
\be
\sum_{n=n_s}^{n_l} E(n) > T.
\ee
We then have a bin of $[n_s,n_l]$ with corresponding size $(n_l-n_s+1)$.
The average value associated with this bin is
\be
\frac{\sum_{n=n_s}^{n_l} Q(n)}{n_l-n_s+1} .
\ee
This procedure is repeated until the data is exhausted. For this method,
the data may be graphically represented either as a single point per bin
(as in the data threshold method above), or as a point (showing the
associated average value) for each measured (non-zero) data point $Q(n)$.

The data threshold method requires no {\em a priori} knowledge, and is a
good predictor of the underlying distribution. However, when there are
few data points, the template threshold method is more reliable. For both
methods, a range of $T$ should be tried and the best $T$ (neither
over- or under-binning) chosen.

\end{multicols}
\end{document}